\documentclass[10pt,aps,twocolumn,prl,showpacs,superscriptaddress,floatfix]{revtex4-2}
\usepackage{amssymb,amsmath,amsfonts}
\usepackage{epsfig,graphicx}
\usepackage{braket}
\usepackage[dvipsnames]{xcolor}
\usepackage{soul}
\usepackage{hyperref}
\hypersetup{colorlinks,citecolor=PineGreen,filecolor=black,linkcolor=black,urlcolor=black}

\usepackage{color}

\begin{document}
\title{Microwave-free vector magnetometry and crystal orientation determination with Nitrogen-Vacancy centers using Bayesian inference}

\author{H. Espin{\'o}s}
\email{hilarioespinos@gmail.com}
\affiliation{Department of Physics, Universidad Carlos III de Madrid, Avda. de la Universidad 30, Legan\'es, 28911  Madrid, Spain} 

\author{O. Dhungel} 
\affiliation{Helmholtz-Institute, GSI Helmholtzzentrum fur Schwerionenforschung, Mainz 55128, Germany}
\affiliation{Johannes Gutenberg-Universit{\"a}t Mainz, 55128 Mainz, Germany}

\author{A. Wickenbrock}
\affiliation{Helmholtz-Institute, GSI Helmholtzzentrum fur Schwerionenforschung, Mainz 55128, Germany}
\affiliation{Johannes Gutenberg-Universit{\"a}t Mainz, 55128 Mainz, Germany}

\author{D. Budker} 
\affiliation{Helmholtz-Institute, GSI Helmholtzzentrum fur Schwerionenforschung, Mainz 55128, Germany}
\affiliation{Johannes Gutenberg-Universit{\"a}t Mainz, 55128 Mainz, Germany}

\affiliation{Department of Physics, University of California, Berkeley, USA}

\author{R. Puebla}
\affiliation{Department of Physics, Universidad Carlos III de Madrid, Avda. de la Universidad 30, Legan\'es, 28911  Madrid, Spain}

\author{E. Torrontegui}
\email{eriktorrontegui@gmail.com}
\affiliation{Department of Physics, Universidad Carlos III de Madrid, Avda. de la Universidad 30, Legan\'es, 28911  Madrid, Spain}

\begin{abstract}
    Nitrogen-vacancy (NV) centers in diamond provide a solid-state platform for quantum sensing. While optically detected magnetic resonance techniques offer high sensitivity, their reliance on microwaves introduces heating and stray electromagnetic fields that can perturb nearby samples. Optical approaches based on cross-relaxation between differently oriented NV centers remove this constraint but have so far required stringent alignment of the external field with crystallographic axes, restricting their practicality. Here we introduce a general framework for microwave-free vector magnetometry at near-zero field that leverages Bayesian inference to extract both the magnetic field vector and the NV orientation directly from photoluminescence maps. An analytical model of cross-relaxation resonances enables efficient inference under arbitrary field and orientation configurations, while naturally incorporating the discrete degeneracies of the NV symmetry. We experimentally demonstrate robust orientation determination and vector-field reconstruction, establishing a general route toward compact and alignment-free NV magnetometers for practical sensing applications.
\end{abstract}

\maketitle

\section{Introduction}
Nitrogen-vacancy centers in diamond are versatile solid-state quantum sensors that exploit optically addressable spin states with long coherence times under ambient conditions. This unique combination has enabled their use as nanoscale thermometers~\cite{Kucsko2013}, electric field sensors~\cite{Dolde2011}, and pressure gauges~\cite{Paudel2024}, while their most widely explored modality is magnetometry~\cite{Maze2008}. Owing to the strong coupling between the NV electronic spin and external magnetic fields, NV-based magnetometers achieve high sensitivities and spatial resolutions down to the nanoscale~\cite{Taylor2008,Dumeige2013,Wolf2015,Chatzidrosos2017,Xie2021}. This capability has led to a broad range of applications, including the imaging of biological cells~\cite{LeSage2013,Glenn2015}, mapping of current distributions in conductors~\cite{Steinert2010,Pham2011}, visualization of magnetic textures in condensed-matter systems such as skyrmions and superconducting vortices~\cite{Balasubramanian2008,Tetienne2014}, and characterization of magnetic devices~\cite{Simpson2016}. NV magnetometry has also enabled nanoscale nuclear magnetic resonance spectroscopy~\cite{Holzgrafe2020}, providing access to molecular-scale spin environments with unprecedented spatial resolution. In many sensing scenarios, complete characterization of the magnetic field requires knowledge of both its magnitude and spatial direction. This has motivated the development of NV-based vector magnetometry~\cite{Rondin2013}, where measurements along the four crystallographic axes of the defect allow full reconstruction of the three-dimensional magnetic field.

Most NV-based magnetometry techniques employ optically detected magnetic resonance (ODMR), where microwave fields (MW) are used to manipulate the electronic spin states of the NV ~\cite{Chipaux2014,Wang2015,Schloss2018,Zhang2018,Weggler2020}. While ODMR provides robust and sensitive readout, the need for MW delivery introduces practical limitations: MW fields can perturb the environment, require galvanic connections that complicate integration, and cause heating in the microwave circuitry that is poses challenges for operation under cryogenic conditions. In contrast, MW-free schemes simplify the experimental architecture, making them better suited for compact and industrial implementations~\cite{Chatzidrosos2021, Wunderlich2021}. One class of optical methods exploits ground-state-level anticrossing, although these approaches require stable bias magnetic fields aligned with the NV crystallographic axis with high precision~\cite{Wickenbrock2016,Zheng2020}. Another class relies on magnetic field-induced mixing of the NV spin sublevels, which alters the spin-dependent optical cycling and produces measurable changes in the intensity of photoluminescence (PL) under optical excitation~\cite{Wunderlich2021,Horsthemke2024b,Horsthemke2024}. In both ODMR-based and MW-free modalities, accurate knowledge of the NV axis orientation is essential for reconstructing the magnetic field vector. Typically, NV centers are located using a confocal microscope, followed by a calibration procedure to determine their crystallographic orientation. A conventional approach relies on ODMR spectroscopy under controlled magnetic fields generated by three-dimensional electromagnets. By fitting the resonance spectra to the NV spin Hamiltonian, the NV orientation can be extracted experimentally~\cite{ChenGB2020,Fukushige2020,Igarashi2020,Wang2025, 11018638}. Alternatively, MW-free methods infer the NV orientation from the angular dependence of optical excitation, using either polarized excitation or structured-beam illumination to map PL patterns~\cite{Alegre2007,Dolan14,ChenB2020}. While these approaches eliminate the need for microwave driving, they generally require high-resolution optical instrumentation and are ultimately limited by the precision of imaging and dipole-modeling techniques.

Recent studies have demonstrated that interactions between NV centers of different crystallographic orientations can be harnessed for magnetometry~\cite{Choi2017,Pellet-Mary2023}. When spin transitions of inequivalent orientations become degenerate, cross-relaxation processes give rise to sharp photoluminescence features at well-defined resonance conditions. These signatures provide an intrinsic, optical contrast mechanism that encodes vectorial information about the external magnetic field. DC magnetometry based on NV–NV cross-relaxation has been demonstrated by tracking the positions of cross-relaxation resonances in fluorescence scans~\cite{Akhmedzhanov2019}, with projected sensitivities in the tens of pT/$\sqrt{\text{Hz}}$~\cite{Akhmedzhanov2017}. The specific protocol proposed in that work, however, relies on prior crystallographic knowledge and precise alignment of the magnetic field to assign resonances unambiguously, which limits its general applicability. Separately, recent work has characterized the corresponding 2D PL maps as a function of control parameters~\cite{Dhungel2024}, but did not address the inverse problem of retrieving field and orientation parameters from such maps.

In this work, we extend cross-relaxation-based magnetometry beyond these alignment-constrained schemes by introducing an alignment-free Bayesian inversion of PL maps. Our framework extracts both the magnetic field vector and the NV orientation directly from RF-free optical measurements. Unlike prior peak-based approaches, our method operates on the full 2D PL response, and remains applicable under arbitrary field configurations and crystallographic orientations, providing full posterior distributions that quantify uncertainties and capture discrete degeneracies inherent to the NV symmetry, while requiring only a single tunable bias-field axis together with a controlled rotation. This generalization establishes a versatile and robust route to RF-free vector magnetometry, broadening the scope of NV-based sensing in regimes where conventional ODMR techniques are impractical.

\section{Results and discussion}
\subsection{NV cross-relaxation features}

\begin{figure*}[t]
    \centering
    \includegraphics[width=1\linewidth]{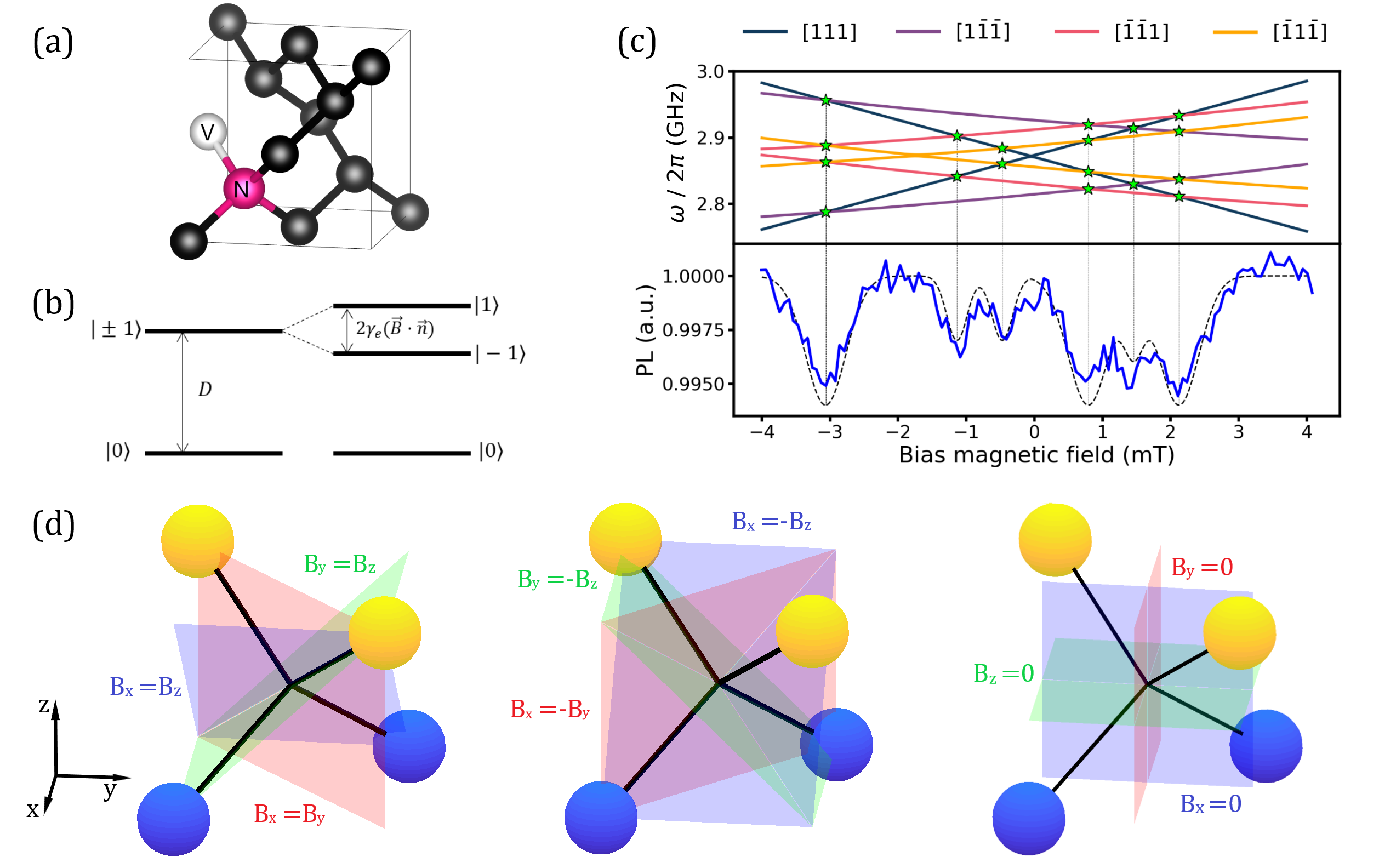}
    \caption{Nitrogen-vacancy (NV) center structure, energy levels, and cross-relaxation resonances. (a) Schematic of a nitrogen-vacancy center in the diamond lattice. (b) Corresponding ground-state energy level structure under a weak magnetic field. (c) Top panel: transition energies between the $m_s=0$ and $m_s=\pm1$ states for NVs along the four crystallographic orientations as a function of a bias magnetic field of variable strength and a fixed transverse magnetic field. Energy crossings between distinct NV classes (green stars) correspond to resonance conditions that enhance cross-relaxation, leading to observable drops in photoluminescence (PL). Bottom panel: experimental (blue solid line) and simulated (black dashed line) PL showing dips aligned with the predicted resonance points. (d) Geometric visualization of the nine resonance planes in magnetic field space. Left and center: symmetry planes corresponding to pairwise resonance conditions between NV orientations. Right: anti-symmetry planes where two resonance conditions are satisfied simultaneously.}
    \label{fig:EnergyCrossings_Symmetry}
\end{figure*}

The negatively charged Nitrogen-Vacancy center in diamond (NV$^-$, denoted NV throughout) is a point defect with ${C}_{3v}$ symmetry consisting of a substitutional nitrogen adjacent to a lattice vacancy, as shown schematically in Fig.\,\ref{fig:EnergyCrossings_Symmetry}(a). In its electronic ground state, the NV center is an electronic spin-1 system, whose Hamiltonian in the absence of strain and electric fields is given by \cite{Doherty2013}
\begin{equation}\label{eq:Hamiltonian}
    H = DS_z^2 + \gamma_e \textbf{B}\cdot \textbf{S}\,,
\end{equation}
where $D = 2\pi\times 2.87$ GHz is the zero-field splitting (ZFS), $\gamma_e=2\pi\times 2.8$ MHz/G is the electron gyromagnetic ratio, $\textbf{B}$ is an applied magnetic field, and $\textbf{S}=(S_x,S_y,S_z)$ are the spin-1 operators. The ZFS term lifts the degeneracy between the $m_s=0$ and $m_s\pm1$ states, creating an energy gap $D$ at zero field, while the Zeeman term splits the $m_s=\pm1$ spin levels. The resulting level structure is shown in Fig.\,\ref{fig:EnergyCrossings_Symmetry}(b). Additional terms such as strain-induced splitting and hyperfine coupling to the host $^{14}$N (or $^{15}$N) nuclear spin are neglected here. In typical ensembles, these contributions vary across the detection volume, so that when integrating the optical signal they manifest mainly as inhomogeneous broadening rather than as a systematic shift of the magnetic-field-dependent response. Uniform, orientation-dependent strain terms could in principle produce systematic shifts, but these are not expected to dominate for the samples and conditions studied.

For an orientation $i$ with unit vector $\mathbf{n_\textit{i}}$, the magnetic field in the NV frame can be decomposed as
\begin{equation}
    B_\parallel^{(i)}=\mathbf{B}\cdot\mathbf{n}_i\,,\hspace{6mm}\mathbf{B}_\perp^{(i)} = \mathbf{B}-B_\parallel^{(i)}\mathbf{n} _i \,.
\end{equation}
Because the ZFS term of the NV Hamiltonian is axially symmetric about its crystallographic axis, the spectrum of Hamiltonian~\eqref{eq:Hamiltonian} depends on the pair $(B_\parallel^{(i)},B_\perp^{(i)})$, but not on the azimuthal direction of $\mathbf{B}_\perp^{(i)}$; any in-plane rotation of the magnetic field leaves the eigenenergies invariant. Moreover, due to the symmetry of the Zeeman term under field inversion, reversing the magnetic field direction, $\mathbf{B}\!\rightarrow\!-\mathbf{B}$, simply interchanges the $\ket{m_s=+1}$ and $\ket{m_s=-1}$ states without altering the energy spectrum. Consequently, for a fixed laboratory magnetic field, if two inequivalent orientations $i$ and $j$ satisfy 
\begin{equation}\label{eq:mag_field_condition}
    |\textbf{B}\cdot\textbf{n}_i|=|\textbf{B}\cdot\textbf{n}_j| \hspace{4mm}\text{for}\hspace{4mm} i\neq j\,,
\end{equation}
then the two Hamiltonians $H_i$ and $H_j$ have identical transition energies from $\ket{m_s=0}$ to $\ket{m_s=\pm1}$. This causes an enhancement of flip-flop processes between the differently oriented NV centers via dipole-dipole coupling. According to the fluctuator model \cite{Choi2017, Pellet-Mary2023}, a subset of rapidly depolarizing NV centers can act as a relaxation channel for the ensemble, mediating spin exchange and accelerating depolarization. 
This results in a population redistribution from the bright $m_s=0$ state to the darker $m_s=\pm1$ manifold, decreasing overall photoluminescence when two groups of inequivalently orientated NV centers are in resonance. 
Equation~\eqref{eq:mag_field_condition} thus defines a set of resonance conditions in magnetic field space. In practice, small transverse zero-field splitting produced by strain and hyperfine structure slightly break perfect axial isotropy; in our regime these act primarily to broaden features and set the linewidths without shifting the resonance conditions.

Figure \ref{fig:EnergyCrossings_Symmetry}(c) shows the transition energies of NVs oriented along four different axes in the presence of a variable magnetic field applied along the [111] direction and a fixed transverse field of 2.2 mT. Crossings between transition energies correspond to cross-relaxation resonances, which manifest as sharp features in the photoluminescence signal. The bottom panel of the figure presents the corresponding experimental data, where these resonances are clearly visible as drops in the photoluminescence intensity. 

Due to the fixed tetrahedral geometry of the diamond, the resonances conditions occur on planar surfaces defined by linear relations between the components of \textbf{B} in the sample reference frame. There are nine such distinct conditions: six symmetry planes
\begin{equation}\label{eq:symmetry_conds}
    B_x^\text{s}=\pm B_y^\text{s}, \hspace{3mm}B_x^\text{s}=\pm B_z^\text{s},\hspace{3mm} B_y^\text{s} = \pm B_z^\text{s}\,,
\end{equation}
corresponding to resonance between a single pair of NV orientations; and three anti-symmetry planes
\begin{equation}\label{eq:antisymmetry_conds}
    B_x^\text{s}=0, \hspace{3mm}B_y^\text{s}=0, \hspace{3mm}B_z^\text{s}=0,
\end{equation}
where two independent pairs become degenerate simultaneously. Here, the superscript `$s$' denotes that the magnetic field components are expressed in the sample (diamond crystal) reference frame. A geometric representation of these nine planes, together with the NV axes, is shown in Fig.\,\ref{fig:EnergyCrossings_Symmetry}(d). Each plane gives rise to a reduction in photoluminescence because of enhanced cross-relaxation between orientation pairs. The anti-symmetry planes, where two such resonances occur simultaneously, produce a larger overall PL contrast.

In the general case, a diamond sample subjected to an arbitrary external magnetic field will not satisfy the resonance conditions given by Eqs.~\eqref{eq:symmetry_conds} and \eqref{eq:antisymmetry_conds}. However, by introducing a tunable bias magnetic field and by rotating the diamond around a fixed axis, the projection of \textbf{B} onto the NV axes can be systematically varied. This makes it possible to bring two NV orientations into resonance, thereby inducing cross-relaxation and a measurable drop in photoluminescence. As shown in Ref.~\cite{Dhungel2024}, by scanning over these external parameters (magnitude of bias magnetic field and rotation angle around a fixed axis), one obtains a two-dimensional photoluminescence map containing sharp features at the resonance conditions, which encode information about the underlying magnetic field or the crystal orientation.

The four NV axes in the diamond lattice form the vertices a regular tetrahedron. As a result, rotations that permute the NV axes without changing the overall geometry leave the system unchanged. These rotations form the proper tetrahedral group $T_d$, which contains 12 operations. In addition, because of the symmetry of  the Zeeman shifts, the system is also invariant under simultaneous inversion of the 4 NV axes, i.e., $\textbf{n}_i\mapsto - \textbf{n}_i$. This extends the effective symmetry group to the proper octaedral group $O$, which include 24 symmetry operations in total. These include eight $\pm2\pi/3$ rotations around the four NV axes, nine rotations by $\pi/2$, $\pi$ and $3\pi/2$ around the Cartesian axes, six $\pi$-rotations around the $[110]$-type directions, and the identity. Under all such transformations, the resonance conditions given by Eq.\,\eqref{eq:mag_field_condition} are preserved, and thus the resulting photoluminescence map remains unchanged. As a result, multiple physically distinct physical orientations of the crystal can yield identical photoluminescence patterns, and this degeneracy must be accounted for in any inversion procedure aimed at recovering magnetic field vectors or crystal orientations.

\subsection{Bayesian framework for parameter estimation}
The proposed experimental method consists in applying a bias magnetic field along a chosen direction and rotating the diamond sample around the axis defined by that same direction. In this section, we formulate a phenomenological model for the expected photoluminescence as a function of two experimental parameters: the rotation angle $\phi$ and the bias magnetic field strength $B^\text{bias}$ applied along the rotation axis. Without loss of generality, we take this axis to be the $z$-axis of the laboratory frame. In addition to the tunable bias field, we include a static external magnetic field $\mathbf{B^{ext}}$, assumed to be fixed and potentially unknown. The total magnetic field is given in the laboratory frame by
\begin{equation}\label{eq:magnetic_field}
    \mathbf{B^\text{lab}}=\mathbf{B^{ext}} + B^\text{bias}\mathbf{u_z},
\end{equation}
where $\mathbf{u_z}$ is the unit vector along the z-axis.
The microwave-free photoluminescence signal acquires magnetic field dependence due to resonant cross-relaxation processes between differently orientated NVs. The crystallographic axes of the diamond are generally misaligned with respect to the laboratory frame, and therefore it is convenient to transform the magnetic field vector at each rotation angle 
$\phi$ into the sample frame, where its projections onto the NV orientations can be directly evaluated. This transformation is expressed as
\begin{equation}\label{eq:magnetic_field_sample}
    \mathbf{B^\text{s}}=O_\text{lab}^\text{s}\cdot  R_z(-\phi)\cdot\mathbf{B}^\text{lab},
\end{equation}
where $R_z(-\phi)$ is the matrix corresponding to a rotation around the laboratory $z$-axis by an angle $\phi$, and $O_\text{lab}^\text{s}$ is the fixed rotation matrix mapping vectors from the laboratory frame to the sample frame. The negative sign indicates that, from the perspective of the diamond frame, the magnetic field appears to rotate in the opposite direction to the physical rotation of the sample in the laboratory frame.

The explicit form of $\mathbf{B}^\text{s}$ depends both on the external magnetic field vector $\mathbf{B}^\text{ext}$ and the relative orientation between the sample and laboratory frames. Its closed-form expression in terms of the control parameters $(B^\text{bias},\phi)$, the external magnetic field and the rotation parameters is given in Supplementary Note 1. Importantly, all features of the PL map, specifically the locations of photoluminescence dips, are determined entirely by the components of $\mathbf{B}^\text{s}$ through the resonance conditions defined in Eqs.~\eqref{eq:symmetry_conds} and \eqref{eq:antisymmetry_conds}. To capture this behavior, we model the PL signal with an expression that explicitly incorporates these resonance conditions,
\begin{equation}\label{eq:photoluminescence}
    \text{PL}(B^\text{bias},\phi)=1-C\sum_iw_iL\left[\delta_i(B^\text{bias},\phi);\Gamma\right],
\end{equation}
where $\delta_i(B^\text{bias},\phi)$ denotes the deviation from the 
$i$-th resonance condition, expressed in terms of the magnetic field components in the sample frame, i.e., $\delta =\{B_x^s-B_y^s,\ B_x^s- B_z^s,\ B_y^s- B_z^s,\  B_x^s+ B_y^s,\ B_x^s+ B_z^s,\ B_y^s+ B_z^s,\ 2B_x^s,\ 2B_y^s,\ 2B_z^s\}$(see Supplementary Note 1 for a explicit closed-form expression of one of these resonance conditions); $L(\delta_i;\Gamma)$ is a symmetric lineshape function centered at $\delta_i=0$ with linewidth $\Gamma$ modeling the PL reduction due to cross-relaxation; $C$ sets the contrast of a single PL dip; and $w_i$ is the relative weight assigned to each resonance condition. In the ideal case of a spatially homogeneous ensemble with equal NV orientation populations being identically illuminated,  $w_i=1$ for single-pair (symmetry plane) crossings and $w_i=2$ for simultaneous (anti-symmetry plane) crossings. Deviations from the ideal situation lead to orientation-dependent weights that modify the effective contrast of each PL dip.

To extract relevant information from the photoluminescence maps, such as the external magnetic field vector or the orientation of the diamond frame, an inverse model is required. This inverse problem is nonlinear, as shown by Eq.\,\eqref{eq:photoluminescence}, and it is subject to experimental noise and model uncertainties. Bayesian estimation~\cite{vandeSchoot2021} provides a natural framework for addressing these challenges: it allows the incorporation of prior knowledge, and yields a full posterior distribution over the parameters of interest. This approach not only identifies the most probable values consistent with the observed PL data but also quantifies confidence and ambiguity arising from measurement noise. 
\begin{figure*}[ht]
    \centering
     \includegraphics[width=0.95\linewidth]{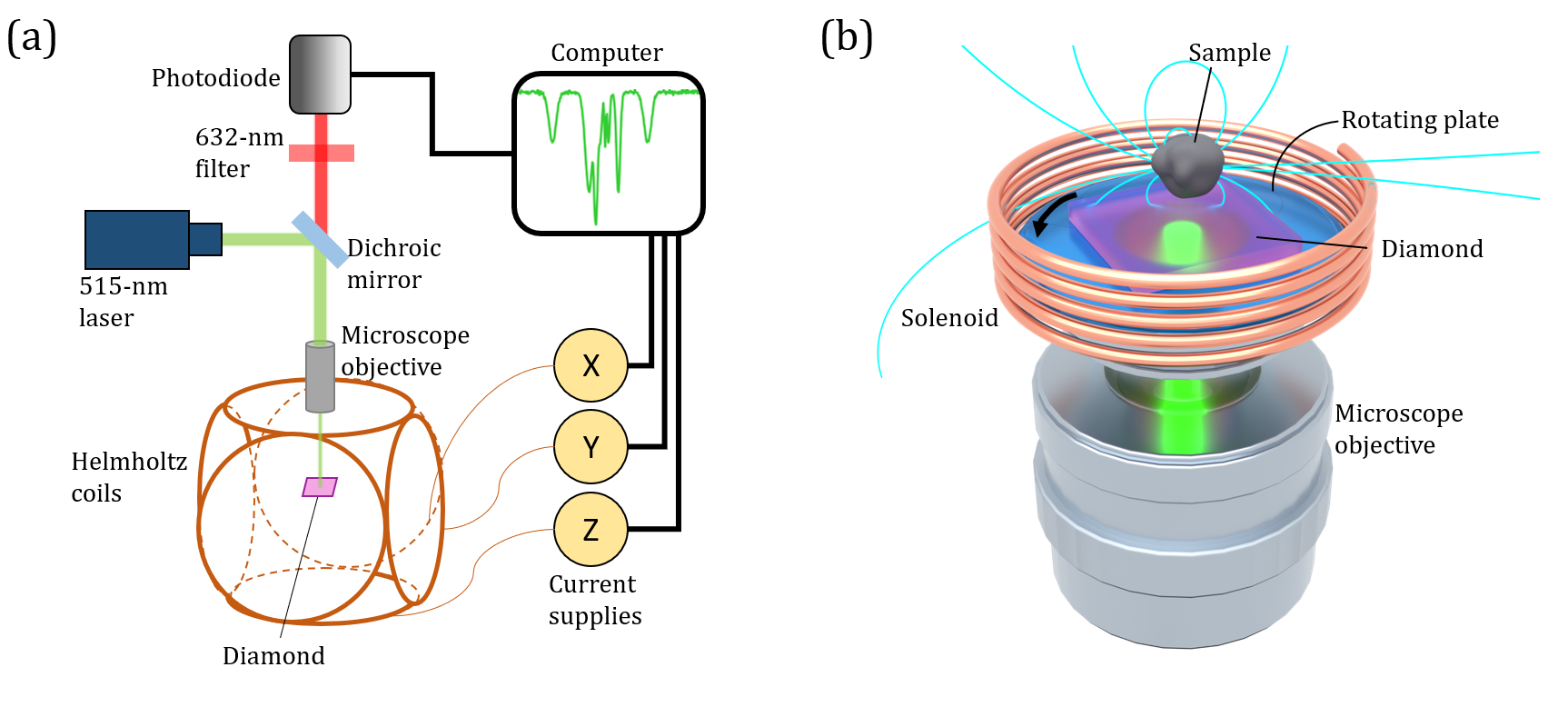}
    \caption{Experimental setup and device concept. (a) Schematic diagram of the experimental setup. (b) Conceptual illustration of the intended device-level implementation for magnetometry.}
    \label{fig:setup}
\end{figure*}

Bayesian estimation is grounded in Bayes’ theorem, which relates the posterior probability of the parameters to the likelihood of the observed data and the prior probability of the parameters,
\begin{equation}
    P(\boldsymbol{\theta}|\mathbf{y}) = \frac{P(\mathbf{y}|\boldsymbol{\theta})P(\boldsymbol{\theta})}{P(\mathbf{y})}.
\end{equation}
Here, $\boldsymbol{\theta}$ denotes the set of unknown parameters to be inferred $(\mathbf{B^\text{ext}},\ O^s_\text{lab},\Gamma,w_i)$, $\mathbf{y}=\text{PL}^\text{exp}(B^\text{bias},\phi)$ represents the observed data, $P(\boldsymbol{\theta}|\mathbf{y})$ is the posterior probability distribution of the parameters given the data, $P(\mathbf{y}|\boldsymbol{\theta})$ is the likelihood of observing the data for a given choice of parameters, and $P(\boldsymbol{\theta})$ is the prior probability distribution of $\boldsymbol{\theta}$, encoding any information known about the parameters before considering the data. The denominator $P(\mathbf{y})$ is a normalization constant, ensuring that the posterior probability distribution integrates to one. In practice, the linewidth $\Gamma$ and the contrast weights $w_i$ can either be fixed from independent calibration measurements and treated as known parameters, or incorporated directly into the Bayesian inference as additional unknown parameters. In the former case, their experimentally determined values are held constant when evaluating the likelihood. In the latter, they are appended to the parameter vector $\boldsymbol{\theta}$ and inferred jointly with the field and orientation parameters.

A common computational bottleneck in Bayesian estimation is the repeated evaluation of the likelihood function, which typically involves simulating the expected signal for each point in parameter space and comparing it with the experimental measurement. In this work, we adopt a Gaussian likelihood, corresponding to the assumption that measurement noise is independent, identically distributed, and approximately Gaussian with standard deviation $\sigma_\text{noise}$. Additional uncertainty arising from imperfect knowledge of the independent experimental variables can be incorporated by propagating their variances into the total noise term,
\begin{equation}
    \sigma^2 = \sigma_\text{noise}^2+\sum_i \left(\frac{\partial \textbf{y}}{\partial x_i}\right)^2\sigma_i ^2,
\end{equation}
where $x_i$ denotes the $i$-th independent variable and $\sigma_i^2$ is its variance. In our case, the independent variables are the axial bias field $B^{\mathrm{bias}}$ and the rotation angle $\phi$, with $\sigma_i^2$ determined by their respective experimental uncertainties.
Under these assumptions, the likelihood takes the form
\begin{eqnarray}\label{eq:likelihood}
P(\mathbf{y}|\boldsymbol{\theta})=\text{exp}\left[-\frac{1}{\sigma^2}\sum_{(B^\text{bias},\phi)}\left(\text{PL}^\text{exp}(B^\text{bias},\phi)\right.\right.\nonumber\\
\left.\left.-\text{PL}^\text{model}(B^\text{bias},\phi|\boldsymbol{\theta})\right)^2\right],
\end{eqnarray}
where the sum runs over all sampled points $(B^\text{bias},\phi)$ at which the PL signal is recorded. For each pair $(B^\text{bias},\phi)$, the quantity $\text{PL}^\text{model}(B^\text{bias},\phi|\boldsymbol{\theta})$ is evaluated using the forward model defined in Eq.~\eqref{eq:photoluminescence}.

Bayesian estimation is particularly well-suited to this problem because it naturally captures the discrete degeneracies arising from the NV symmetry, representing them as distinct peaks in the posterior distribution rather than forcing a single, potentially misleading, best-fit solution. Another key advantage of our forward model used to simulate PL maps is that it avoids repeated Hamiltonian diagonalization during likelihood evaluation. By expressing the photoluminescence signal analytically in terms of geometric resonance conditions in the magnetic field space, we derive a closed-form model for the expected measurement $\text{PL}^\text{model}(\boldsymbol{\theta})$. This yields a substantial computational speedup, enabling fast and scalable Bayesian inference without compromising accuracy.

From the posterior distribution $P(\boldsymbol{\theta}|\mathbf{y})$, one can compute marginal probability distributions for individual parameters by integrating out the others. These marginal distributions capture the individual confidence intervals and correlations between parameters and are essential for interpreting the results of the inference.

\subsection{Experimental implementation}
The fluorescence measurements were performed with a wide-field fluorescence microscope, as shown in Fig.~\ref{fig:setup}(a). Excitation is provided by the 515\,nm Toptica iBeam smart laser. The laser beam is focused on the diamond using 50× Mitutoyo long-working-distance objective (NA = 0.45). The PL is collected through the same objective and detected with a photodiode. An arbitrarily directed magnetic field is generated with a set of Helmholtz coils around the sample. The magnetic fields were calibrated using ODMR (used here as a convenient, but not required, calibration method). The diamond used in all experiments is an (111)-oriented, HPHT-grown single crystal. It has an estimated substitutional nitrogen concentration of $\sim$100 ppm and was electron-irradiated (energy 3 MeV, dose 10$^{18}$ cm$^{-2}$), followed by annealing at 1050$^{\circ}\mathrm{C}$ for 2h, yielding an NV ensemble with an estimated concentration $>5$\,ppm. However, the method is applicable to other samples, including CVD-grown and arbitrarily cut diamonds, provided that cross-relaxation features are measurable; such features have been observed in ensemble samples down to $\sim$0.3 ppm NV density~\cite{Zheng2025}.

In Fig.~\ref{fig:setup}(b) we illustrate the intended device-level implementation. The diamond is mounted on a rotating plate and mechanically rotated about the laboratory $z$-axis while an axial bias field is applied along the same axis, such that the acquired PL map encodes the external-field vector through its dependence on $B^\text{bias}$ and the rotation angle $\phi$.

\subsection{Application to orientation determination}\label{section:orientation}
\begin{figure}[t!]
    \centering
     \includegraphics[width=1\linewidth]{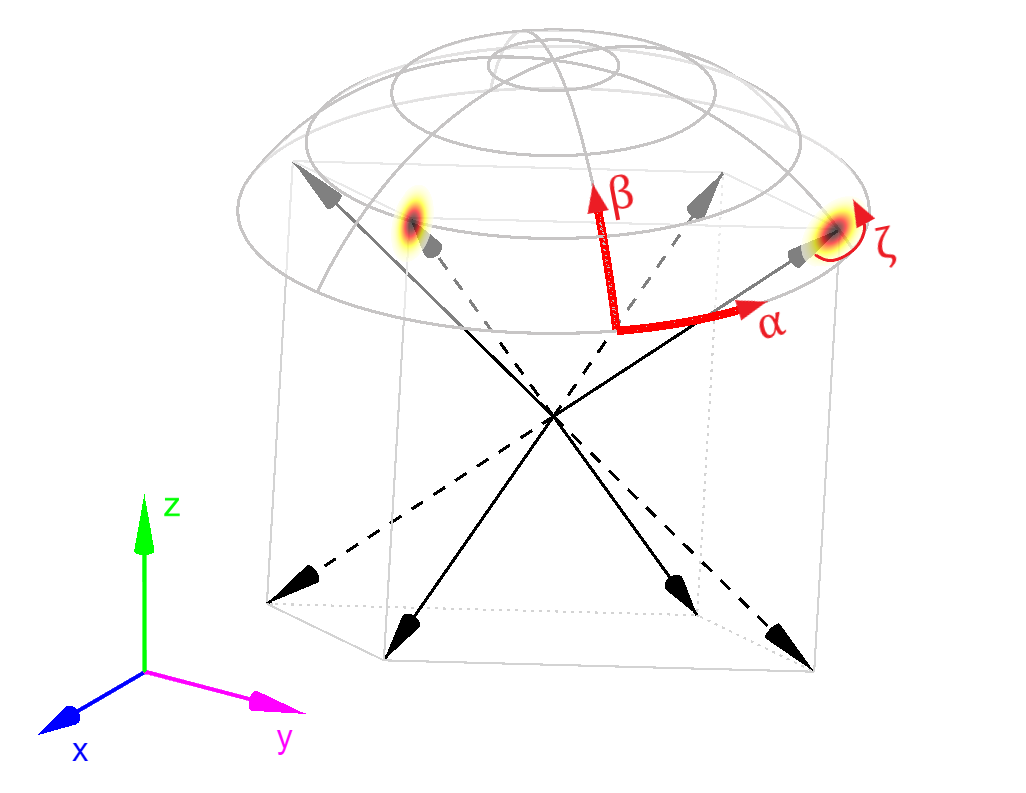}
    \caption{Schematic representation of the restricted orientation space used in the Bayesian inference.  Red axes indicate the Euler angles $(\alpha,\beta,\zeta)$ defining the rotation of the NV frame with respect to the laboratory coordinates ($x,y,z$). To exploit the tetrahedral symmetry of the nitrogen-vacancy (NV) center configuration, the exploration domain is limited to $\alpha\in[0,2\pi]$, $\beta\in[0,\theta_c],$ and $\zeta\in[0,2\pi/3]$, where $\theta_c=\arccos(1/\sqrt{3})$. Within this region, two inversion-related NV orientations ($\mathbf{n}_i$ in solid lines and $-\mathbf{n}_i$ ) remain, highlighted as localized probability distributions on the sphere.}
    \label{fig:orientations}
\end{figure}

To determine the orientation of a diamond crystal from its photoluminescence, we can apply a fixed transverse field $\mathbf{B^{ext}}=b_\perp \mathbf{u_x}$ together with a variable axial bias field $B^\text{bias}$ along $z$. In the envisioned protocol, the sample would be physically rotated around the laboratory $z$-axis to generate a two-dimensional PL map. In our proof-of-concept implementation, we instead rotate the transverse magnetic field using two additional orthogonal pairs of Helmholtz coils (see Fig.~\ref{fig:setup}(a)). This implementation is experimentally more convenient in our current setup and produces the same relative rotation between the crystal frame and the applied field; it requires electronic control of the transverse field in the $xy$ plane, but is equivalent, at the level of the model, to keeping the field fixed and rotating the sample.

\begin{figure*}[t!]
    \centering
     \includegraphics[width=0.99\linewidth]{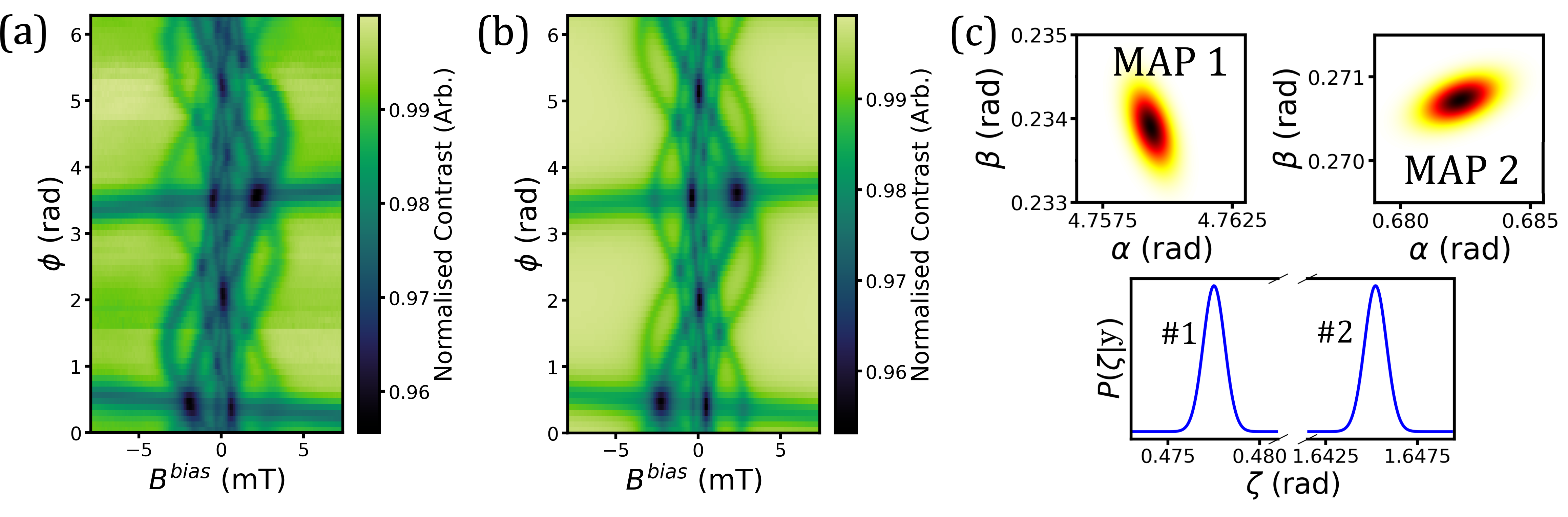}
    \caption{Orientation determination from photoluminescence (PL) maps. (a) Experimental photoluminescence map as a function of the axial bias field $B_\text{bias}$ and the transverse field angle $\phi$, for a diamond crystal with a randomly selected orientation. The PL contrast arises from cross-relaxation between differently oriented nitrogen-vacancy centers at specific resonance conditions.  (b) Simulated PL map with the inferred parameters. (c) Marginal posterior distributions for the diamond orientation obtained from the Bayesian analysis of the experimental data. The top panels show the two high-probability regions in the ($\alpha$, $\beta$) space, corresponding to the two inversion-related orientations that produce identical PL maps. The bottom panel displays the marginal posterior over $\zeta$ with peaks associated with the two peaks shown above.}
    \label{fig:orientation_determination}
\end{figure*}
The unknown orientation of the diamond with respect to the laboratory frame is described by the rotation matrix $O^{\text{s}}_{\text{lab}}$, which we parameterize as a composition of three successive rotations: a rotation by an angle $\alpha$ about the laboratory $z$ axis, followed by a rotation by an angle $\beta$ about the $[1\bar{1}0 ]$ axis of the diamond crystal, and finally a rotation by an angle $\zeta$ about the $[111]$ axis. In the sample reference frame, the magnetic field is transformed as
\begin{equation}
    O^{\text{s}}_{\text{lab}} = R_{[111]}(-\zeta)\cdot R_{[1\bar10]}(-\beta)\cdot R_z(-\alpha).
\end{equation}
Note that this specific parametrization is chosen for convenience: the first two rotations tilt the $[111]$ crystal direction into the desired orientation, while the final rotation about $[111]$ by $\zeta$ fixes the remaining degree of freedom. Other equivalent choices work equally well. Varying $(\alpha, \beta, \zeta)$ over the full ranges $\alpha \in [0, 2\pi)$, $\beta \in [0, \pi]$, and $\zeta \in [0, 2\pi)$ spans the full group of three-dimensional rotations $\mathrm{SO}(3)$. However, as explained in previous sections, the space of photoluminescence signals is not uniquely mapped by $\mathrm{SO}(3)$ due to the discrete symmetry of the NV system. In the full $(\alpha, \beta, \zeta)$ parameter space, there are 24 equivalent configurations that yield identical PL maps, corresponding to the cardinal elements of the $O$ symmetry group.

To avoid redundant exploration of equivalent orientations, we restrict the parameter space by exploiting symmetry. First, we limit the range of the final rotation $\zeta$ to the interval $[0, 2\pi/3]$, since rotations by $2\pi/3$ around the $ [111 ]$ axis leave the NV orientation structure invariant. This reduces the number of equivalent configurations from 24 to 8. A further reduction is achieved by bounding the polar angle $\beta$ to the interval $[0, \theta_c]$, where $\theta_c = \arccos(1/\sqrt{3})$ is the angle between the $[111]$ and $[001]$ directions. The resulting restricted portion of orientation space is illustrated in Fig.~\ref{fig:orientations}. The solid and dashed lines represent the directions of two inversion-related tetrahedra ($\mathbf{n}_i$ and $-\mathbf{n}_i$)  corresponding to the four crystallographic NV axes. Within the explored domain, one direction from the solid tetrahedron and one from the dashed tetrahedron fall inside the allowed region, highlighted as localized probability distributions. In the general case, at least one of the eight NV directions lies within the explored parameter space, and in most cases two inversion-related directions fall inside this region. This reduction of the orientation space enables efficient and non-redundant Bayesian inference over a minimal, symmetry-inequivalent domain.

To validate the proposed orientation determination method, we analyze experimental photoluminescence data obtained from a diamond crystal with an unknown orientation relative to the laboratory frame. Two perpendicular Helmholtz coils are used to generate a rotating transverse magnetic field. In this configuration, the applied magnetic field consists of $b_\perp=1$ mT transverse field rotating by an angle $-\phi$ in the $(x, y)$ plane, $\mathbf{B^{ext}}=b_\perp\left[\cos(\phi)\mathbf{u_x}-\sin(\phi)\mathbf{u_y} \right]$, which is equivalent to rotating the sample by an angle $\phi$. An additional variable axial component is applied along the laboratory $z$-axis to complete the field configuration. Figure~\ref{fig:orientation_determination}(a) shows the experimental PL map as a function of the transverse field angle $\phi$ and the axial bias field $B^\text{bias}$. The data have been normalized to remove a smooth magnetic-field-dependent background (see Supplementary Note 2). The map exhibits characteristic dips in photoluminescence intensity at specific combinations of $(\phi, B^\text{bias})$ corresponding to resonance conditions between NV orientations. The dataset includes 154 bias-field values and 72 rotation angles. The noise standard deviation used in the likelihood model is $\sigma_\text{noise}=0.0018$ , while the experimental uncertainties are estimated as $\sigma_{B^{\text{bias}}}=1\mu T$ and $\sigma_\phi=1^o$.

\begin{figure*}[t!]
    \centering
    \includegraphics[width=0.95\linewidth]{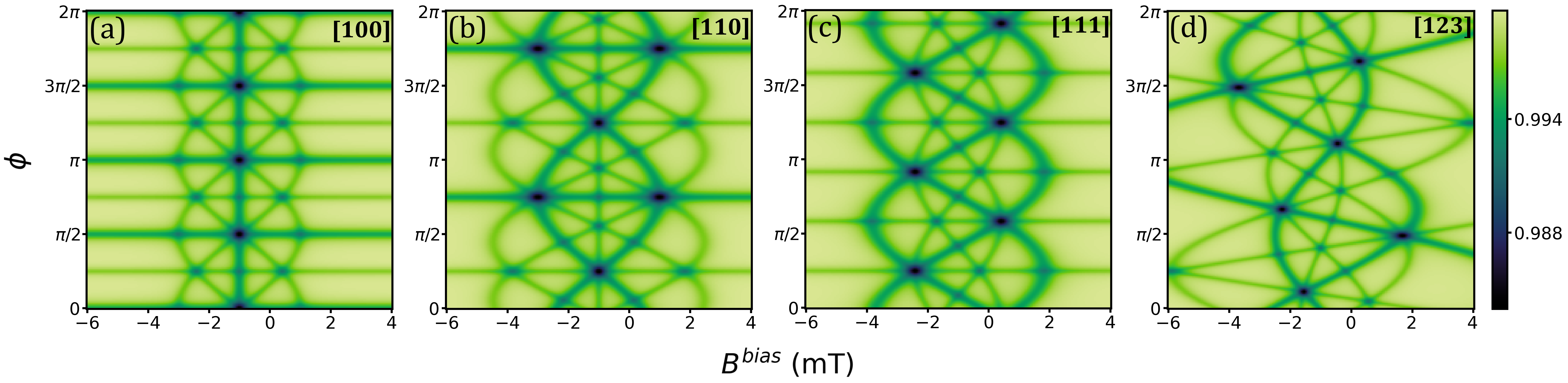}
    \caption{Simulated photoluminescence maps as a function of the rotation angle $\phi$ and the bias field $B^\text{bias}$ for four sample orientations relative to the laboratory frame. In all cases, the external magnetic field is fixed to $\textbf{B}=b_\perp\mathbf{u_x}+b_z\mathbf{u_z}$, with $b_\perp=2$ mT and $b_z=1$ mT. The sample is oriented such that the laboratory $z$-axis (around which the sample is rotated) is aligned with (a) the [100] crystal axis, (b) [110], (c) [111], (d) an arbitrary non-symmetric direction, chosen here as [123], outside the symmetry planes shown in Fig.~\ref{fig:EnergyCrossings_Symmetry}(d).}
    \label{fig:magnetometry}
\end{figure*}
Bayesian inference is then used to recover the rotation parameters that define the orientation of the diamond frame relative to the laboratory frame. Figure \ref{fig:orientation_determination}(b) displays the simulated PL map obtained using the analytical model of Eq.~\eqref{eq:photoluminescence} with the parameters extracted from the Bayesian analysis, showing good qualitative agreement with experiment. In this analysis, the linewidth 
$\Gamma$ and the overall contrast parameters are obtained from an independent calibration subset and are treated as fixed inputs in the Bayesian inference; they could also be included as additional inferred parameters, at the cost of increased computational runtime. Residual systematic deviations between experimental and simulated signals are consistent with a slight miscalibration of the transverse rotating-field amplitude, which primarily rescales the sinusoidal modulation and can therefore yield the observed mildly ``stretched" cross-relaxation features when comparing experimental and simulated PL maps. The Bayesian inference spans the reduced, symmetry-inequivalent domain of the rotation parameters $(\alpha,\beta,\zeta)$. The resulting posterior distribution, shown in Fig. \ref{fig:orientation_determination}(c), reveals two well-localized high-probability regions related by inversion symmetry. The upper panels show $\alpha$-$\beta$ marginal posteriors around each maximum a posteriori (MAP) estimate. The corresponding one-dimensional marginal over $\zeta$ is shown below. The MAP estimates are
$\alpha_1 = 4.7587(8)$ rad, $\beta_1 = 0.2342(3)$ rad, and $\zeta_1 = 0.4775(7)$ rad for the first solution,
and
$\alpha_2 = 0.6832(9)$ rad, $\beta_2 = 0.2705(3)$ rad, $\zeta_2 =1.6452(7)$ rad for the symmetry-related counterpart. Both parameter sets reproduce the experimental PL pattern, confirming the equivalence of the two inversion-related orientations. The analysis demonstrates that the method successfully recovers the diamond orientation from purely optical data, even under noisy conditions. While higher signal-to-noise ratio improves the statistical precision of the reconstructed angles, the accuracy achieved in the present experiment is mainly limited by systematic effects, such as polarization-dependent excitation and illumination inhomogeneity, which will be discussed later in the manuscript.

\subsection{Application to magnetometry}
In the previous application, a well-defined magnetic field was used to infer the diamond orientation; here, that orientation serves as the reference for magnetic-field sensing. With the crystal orientation now calibrated and known, the methodology can be applied in reverse to determine an unknown external magnetic-field vector from photoluminescence data. The experimental protocol remains unchanged: the diamond is rotated around the laboratory $z$-axis while a tunable bias field is applied along the same axis, and the resulting PL map is recorded as a function of the rotation angle and bias-field strength. As in the orientation-determination procedure, the positions of the photoluminescence dips encode the projections of the magnetic field onto the NV axes. By inverting the PL map using the analytical model together with Bayesian inference, the full vector components of the unknown magnetic field in the laboratory frame can be reconstructed.

To define the parameters for inference, we parameterize the external magnetic field as
\begin{equation}
\mathbf{B^{ext}}=b_\perp\left[\cos(\varphi_0)\mathbf{u_x}+\sin(\varphi_0)\mathbf{u_y}\right]+b_z\mathbf{u_z},
\end{equation}
with $b_z$, $b_\perp\ge0$, and $0\le\varphi_0<2\pi$ treated as the free parameters in the Bayesian analysis. Because the bias field is applied along $z$, see Eq.\,\eqref{eq:magnetic_field}, the parameter $b_z$ effectively acts as an offset in the bias field, shifting the entire PL pattern along the horizontal axis by $-b_z$. Similarly, from Eq.\,\eqref{eq:magnetic_field_sample} it follows that the azimuthal angle $\varphi_0$ and the rotation angle $\phi$ enter the transformation in equivalent ways.  
As a result, changing $\varphi_0$ by a given amount simply displaces the PL map along the $\phi$ axis. 

The choice of sample orientation plays a critical role in magnetometry with this method. When the laboratory $z$-axis aligns with a symmetry axis of the NV tetrahedron, the PL map exhibits discrete rotational periodicities and mirror symmetries determined by the order of the stabilizer subgroup associated with that axis. These symmetries lead to ambiguities in the inferred magnetic-field orientation, producing multiple equivalent solutions spaced by the symmetry period. Consequently, the reconstruction of the external magnetic field is not unique unless the symmetry is broken by choosing a different rotation axis or introducing additional measurements. This behavior is illustrated in Fig.\,\ref{fig:magnetometry}, which shows simulated PL maps for four representative cases. In each of them, the cross-relaxation features are produced by an external magnetic field with components $(b_z,b_\perp,\varphi_0)=(1\text{ mT, 2 mT, 0)}$. The only difference between the panels is the choice of laboratory 
$z$-axis: (a) along the [100] crystal direction, (b) along [110], (c) along [111], and (d) along an arbitrary axis not lying in any magnetic-field symmetry plane. The first three configurations display pronounced symmetries. In Fig.\,\ref{fig:magnetometry}(a), the system is invariant under $\pi/2$ rotations, yielding a PL pattern that repeats with period $\pi/2$ in $\phi$, and with mirror symmetries every $\pi/4$. In Fig.\,\ref{fig:magnetometry}(b), the pattern repeats every $\pi$ with mirror symmetries every $\pi/2$. In Fig.\,\ref{fig:magnetometry}(c), the periodicity is $2\pi/3$, with mirror symmetries every $\pi/3$. By contrast, the arbitrary axis case shown in Fig,~\ref{fig:magnetometry}(d) shows no discernible symmetry, thereby eliminating orientation ambiguities. As anticipated, the patterns in all four cases are shifted by $-1$ mT along the $B^\text{bias}$ axis due to the $b_z$ component of the external field. 

To demonstrate magnetic-field reconstruction under realistic conditions, we apply the Bayesian inference framework using the crystal orientation determined in the previous section.  
In this case, the laboratory $z$-axis is not aligned with any of the high-symmetry directions of the NV tetrahedron, which ensures that the photoluminescence map does not exhibit rotational or mirror symmetries.  
This choice of orientation allows a unique reconstruction of the external magnetic field vector from a single measurement configuration. 
\begin{figure}[t]
    \centering
     \includegraphics[width=1\linewidth]{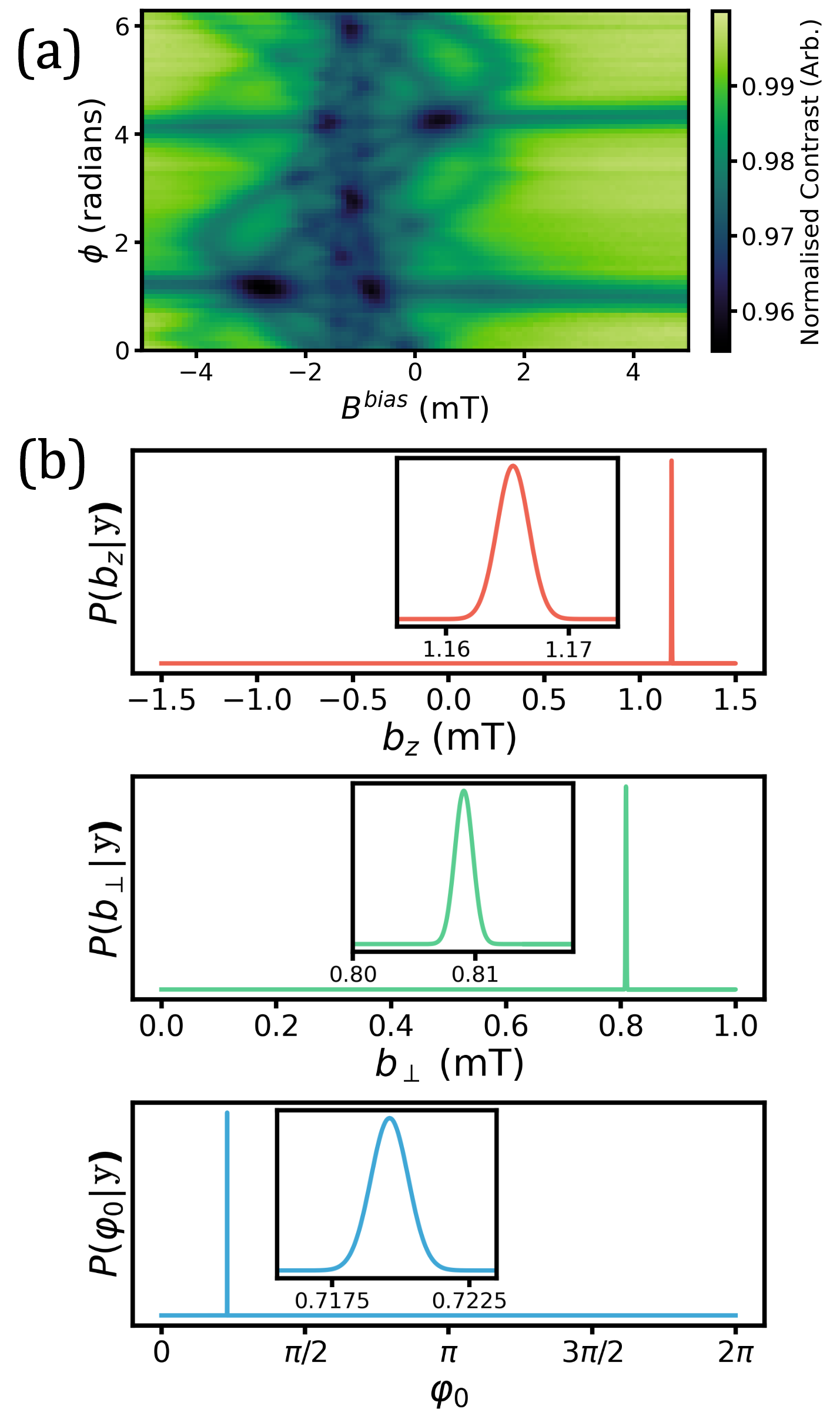}
    \caption{Magnetic-field reconstruction from photoluminescence (PL) data. (a) Measured PL map as a function of the rotation angle $\phi$ and the applied bias magnetic field $B^{\text{bias}}$ for a new magnetic-field configuration. The diamond is kept fixed in the previously determined orientation. (b) Marginal posterior probability distributions obtained from Bayesian inference. Each panel shows the inferred distribution for one of the magnetic-field parameters ($b_z$, $b_\perp$, and $\varphi_0$). Insets provide zooms of the peak regions, from which the maximum a posteriori estimates are extracted.}
    \label{fig:experimental_magnetometry}
\end{figure}

The total external field is described by Eq.\,\eqref{eq:magnetic_field}, with parameters $(b_z,b_\perp,\varphi_0)$ different from those of the calibration procedure.  
A new photoluminescence map was recorded as a function of the bias field $B^{\text{bias}}$ and the rotation angle $\phi$. The measurement conditions and uncertainty parameters used in the Bayesian analysis are identical to those described in the orientation-determination section. The resulting experimental PL map, shown in Fig.\,\ref{fig:experimental_magnetometry}(a), exhibits the same overall pattern as that obtained during orientation determination, but with parameter-dependent modifications: a horizontal displacement associated with the new longitudinal component $b_z$, a vertical phase shift arising from the modified azimuthal angle $\varphi_0$, and a change in the amplitude of the sinusoidal cross-relaxation features reflecting the different transverse field strength $b_\perp$.

We apply Bayesian analysis using the previously inferred orientation matrix $O_\text{lab}^\text{s}$ to find the parameters $(b_z,b_\perp,\varphi_0)$ of the external magnetic field. The results are shown in Fig.\,\ref{fig:experimental_magnetometry}(b). The three marginal distributions reveal well-localized peaks, indicating that they are identifiable within the chosen experimental configuration. The maximum posteriori estimates are $b_z=1.165(2)$\,mT, 
$b_\perp=0.8091(9)$ mT, and $\varphi_0=0.7196(8)$ rad. The resulting values are compatible with independent ODMR measurements, which provides an additional check on the validity of the method. A corresponding simulated map, obtained using the inferred parameters, is presented in Supplementary Note 1.

This example highlights the flexibility of the Bayesian framework: by separating the estimation of the crystal orientation and the magnetic-field parameters, the same optical measurement protocol can be used for both calibration and sensing, enabling robust RF-free vector magnetometry without prior alignment or field biasing. An analysis of the magnetometer's practical dynamic range and the limitations imposed by magnetically induced state mixing is provided in Supplementary Note 3.

\subsection{Analysis of systematic errors}
\begin{figure}[ht]
    \centering
     \includegraphics[width=0.8\linewidth]{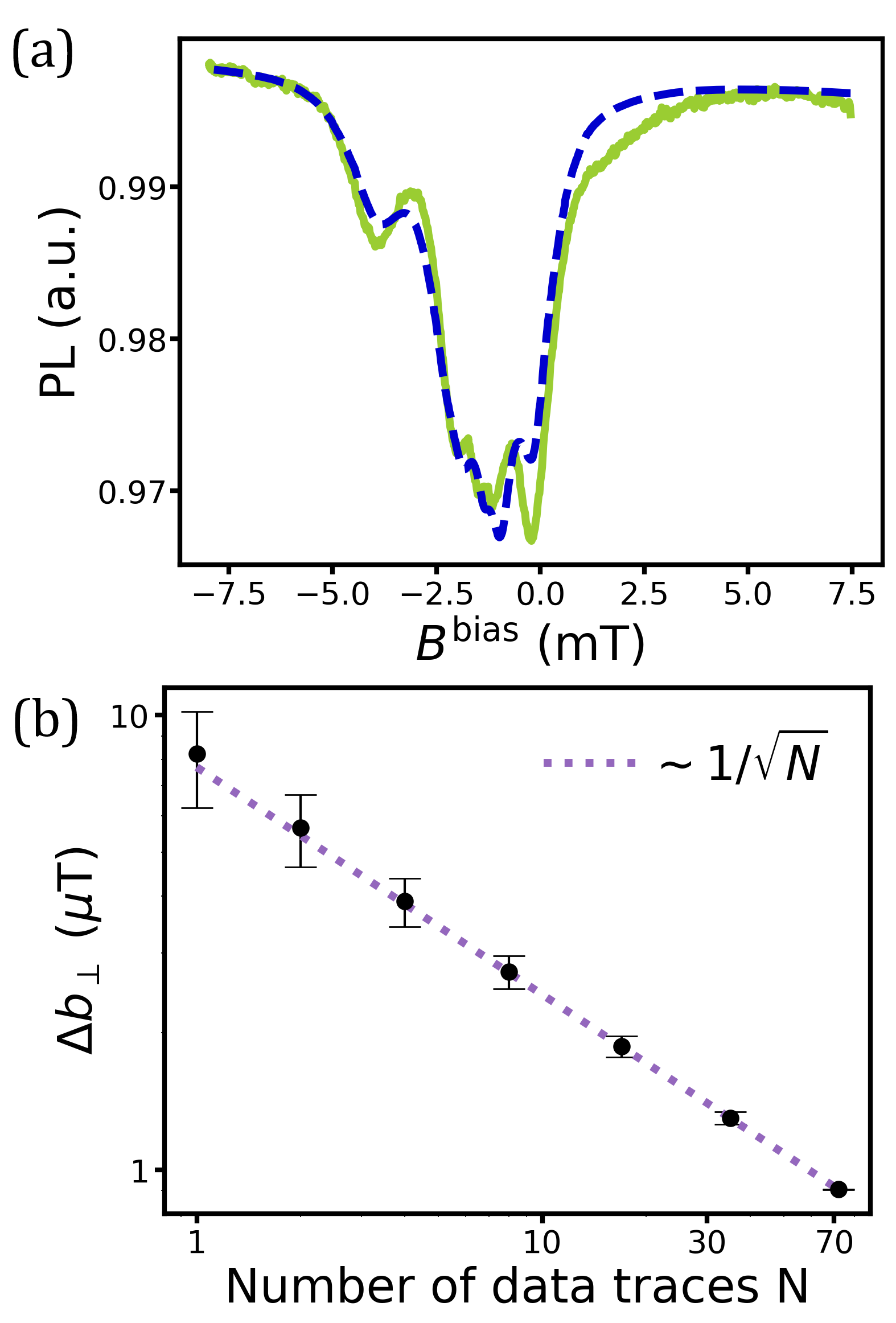}
    \caption{Systematic deviations and scaling of inference uncertainty. (a) Experimental photoluminescence (PL) trace (solid green) and corresponding simulated signal with the analytical model (dashed blue) at a fixed angle of $\phi=335^o$ for the parameter configuration of Fig.~\ref{fig:experimental_magnetometry}. (b) Uncertainty of the inferred transverse magnetic-field component as a function of the number of PL traces used in the Bayesian estimation. Black points show the mean posterior width obtained from 1000 repetitions for each $N$; error bars indicate the standard deviation of $\Delta b_\perp$ across the 1000 repetitions; and the dashed line indicates the expected $1/\sqrt{N}$ scaling for independent measurements.}
    \label{fig:error_analysis}
\end{figure}
The value of $\sigma_\text{noise}$ used in the Bayesian likelihood was fixed to $0.0018$. This value corresponds to the standard deviation of the residuals between the modeled and measured PL signals and reflects the general agreement between the experiment and the model. Although the shot noise of the measurement is one order of magnitude lower with our experimental conditions, the residuals are dominated by systematic effects not captured by the analytical model. This is evident in Fig.~\ref{fig:error_analysis}(a), where one experimental PL trace at a fixed angle (green) and the corresponding calculated signal (blue dashed) agree in the overall structure and in the positions of the cross-relaxation dips, but differ noticeably in the relative contrast of individual features. In addition, the measured trace exhibits a weak shoulder adjacent to the central feature (most visible at negative magnetic fields) that is not reproduced by the analytical model. Similar splittings in microwave-free PL features have been attributed to hyperfine-enabled extra degeneracies and dipolar relaxation channels with other spin species~\cite{Pellet-Mary2021,Pellet-Mary2023,Irena2025,Zheng2025}. In our sample, this structure could arise from hyperfine-induced splittings from strongly coupled naturally abundant $^{13}$C nuclear spins in the diamond. Altogether, these deviations correspond to amplitude differences of order $10^{-3}-10^{-2}$ in the PL intensity.

Among the systematic effects contributing to these discrepancies is the polarization dependence of NV excitation. When linearly polarized light is used, the four families of NVs aligned along different crystallographic axes couple differently to the excitation beams, leading to unequal PL contrasts for each NV family. Using circularly polarized light may partially alleviate this effect by equalizing the excitation efficiency across NV orientations, although complete averaging would require isotropic excitation. A second source of systematic variation arises from inhomogeneous illumination, where the effective laser intensity incident on the sample changes with rotation angle, leading to apparent modulation of the PL signal unrelated to magnetic effects. This effect could, in principle, be compensated by synchronizing the rotation of the laser source with that of the diamond.
Furthermore, mechanical misalignment during rotation can cause small displacements of the diamond relative to the magnetic field center, resulting in the NV ensemble experiencing slightly different field magnitudes or directions throughout the measurement. Such shifts can distort the PL features and bias the inferred parameters if not properly corrected. Temperature and laser power variations can affect the visibility of the cross-relaxation features, primarily through changes in linewidth and contrast, but the crossing positions are largely insensitive to uniform shifts to first order~\cite{Dhungel2024,dhungel2024nd, Zheng2025}. Finally, using isotopically enriched $^{12}$C would eliminate additional spectral structure and broadening. 

The choice of $\sigma_\text{noise}$ directly influences the uncertainties of the inferred parameters: smaller values lead to narrower posterior distributions and thus more precise estimates. In principle, if the experiment were optimized to minimize systematic error, the residuals could approach the shot-noise limit, allowing for a smaller $\sigma$ and improved precision. Under the present conditions, however, $\sigma_\text{noise}=0.0018$  represents a realistic effective uncertainty that balances statistical noise with model mismatch, ensuring that the reported confidence intervals faithfully reflect the experimental limitations.

In addition to the choice of $\sigma_\text{noise}$, the number of experimental data points used in the inference also affects the width of the posterior distributions. To quantify this dependence, we simulated Bayesian estimation using subsets of the data containing $N$ PL traces, each corresponding to a fixed rotation angle. For each value of $N$, the estimation was repeated 1000 times, and the resulting posterior widths were averaged. As shown in Fig.~\ref{fig:error_analysis}(b), the uncertainty of the inferred parameters, illustrated here for $b_\perp$, decrease approximately as $1/\sqrt{N}$, consistent with the scaling expected for independent measurements. In the specific case of magnetometry, the inference can already be performed with a single PL trace ($N=1$), although with significantly larger uncertainty. In contrast, orientation determination requires measurements at multiple angles, since information from a single PL trace is insufficient to uniquely define the three rotational degrees of freedom. This result demonstrates that the Bayesian framework remains robust even with limited data, while the precision of the estimates systematically improves as additional measurement angles are included.

Finally, for robust performance in a deployable device, the sensing volume should be engineered to minimize spatial inhomogeneities that otherwise wash out cross-relaxation structure upon ensemble averaging. In particular, strain or temperature gradients, variations in NV density, or bias-field nonuniformity across the detection region broaden the effective features (captured phenomenologically by a larger 
$\Gamma$ in the PL model). Optimizing the sample and collection volume therefore directly improves reconstruction fidelity and reduces the estimation uncertainties.

\section{Conclusions}
We have presented a microwave-free method for vector magnetometry and crystal orientation determination in NV-diamond systems operating near zero magnetic field. By exploiting cross-relaxation features between NV orientations and modeling the photoluminescence signal through simple geometric resonance conditions, we formulate an analytical description of the photoluminescence map that avoids computationally expensive Hamiltonian diagonalization. This model, combined with Bayesian estimation, enables robust reconstruction of the diamond orientation and the determination of an unknown external magnetic field vector, while also accounting for the discrete symmetries of the NV system.

We validate the method experimentally, demonstrating that the posterior distribution correctly resolves the target parameters. The computational efficiency allows fast inference even with noisy datasets, with uncertainties determined primarily by photoluminescence contrast and feature sharpness, making it suitable for real-time applications. In particular, once the crystal orientation is known and fixed, it is possible to estimate the magnetic-field vector from a single-angle trace rather than a full two-dimensional PL map, lowering both acquisition and computational overhead. For reference, with our current MATLAB implementation based on grid evaluation of the likelihood and 8-core parallelization, a single-angle update runs in $\sim10$s for a $100\times100\times100$ grid, while single-trace acquisition can be performed on a timescale of seconds. Further speedups are expected from optimized implementations, such as compiled code  or GPU, and more efficient Bayesian estimators than exhaustive sweeps. For slowly varying fields, the inference could be sequentially updated by using the posterior at one time step as the prior for the next.

A simultaneous determination of the crystal orientation and the external magnetic field from a single PL map is, in principle, possible up to a global rotation around the laboratory $z$-axis. This ambiguity reflects a fundamental limitation of the method: because the sample is rotated around $z$, any configuration in which both the diamond and the magnetic field are rotated by the same angle about this axis produces an indistinguishable PL map. This ambiguity could be lifted with complementary measurements, like rotating around a second axis or introducing an independent reference field. In practice, the additional challenge in such joint estimations lies in computational cost, as the likelihood evaluation becomes increasingly demanding with the number of free parameters to be inferred.

Finally, our approach could enable self-calibrated (absolute) microwave-free magnetometry. In this scenario, calibration would be performed internally by exploiting additional PL landmarks with field positions set by intrinsic NV physics, such as the GSLAC/ESLAC and other reproducible low-field PL features~\cite{Irena2025,Zheng2025,Pellet-Mary2021}. These internal references provide fixed markers on the magnetic-field axis, allowing the Bayesian inference to jointly estimate the magnetic field and the field-to-control conversion factor without auxiliary sensors.

These results open a route towards compact, alignment-free absolute NV magnetometers that require neither radiofrequency driving nor high-resolution optical imaging, broadening the range of experimental scenarios, where NV sensing can be deployed.

\section{Funding statement}
H.E., R.P., and E.T. acknowledge financial support from the Spanish Government via the projects PID2021-126694NA-C22 and PID2024-161371NB-C21 (MCIU/AEI/FEDER, EU) and project TSI-069100-2023-8 (Perte Chip-NextGenerationEU). H. E. acknowledges the Spanish
Ministry of Science, Innovation and Universities for funding through the FPU program (FPU20/03409). R. P. and E. T. acknowledge the Ram{\'o}n y Cajal RYC2023-044095-I and RYC2020-030060-I research fellowships. O.D., A.W., and D.B. acknowledge the funding by the German Research Foundation (DFG) in the framework to the collaborative research center ``Defects and Defect Engineering in Soft Matter'' (SFB1552) under Project No. 465145163, European Commission’s Horizon Europe Framework Program under the Research and Innovation Action MUQUABIS GA no. 101070546, German Federal Ministry of Education and Research (BMBF) within the Quantumtechnologien program (DIAQNOS, project no. 13N16455) and the project Quantum Sensing for Fundamental Physics (QS4Physics) from the Innovation pool of the research field Helmholtz Matter of the Helmholtz Association.

\section{Acknowledgements}
Not applicable.

\section*{Author contribution}
H.E. developed the analytical model, performed the numerical calculations, and wrote the manuscript. O.D. carried out the experiments. A.W. and D.B. supervised the experimental work and contributed through discussions and interpretation of the results. R.P. and E.T. supervised the theoretical work and contributed to the development of the analysis.  All authors reviewed the final manuscript.

\section*{Competing interests}
The authors declare no competing interests.

\section*{Data availability }
The data that support the findings of this study are openly available in the Zenodo data repository~\cite{zenodo}. These data include experimental measurements and numerical simulation outputs.

\section*{Code availability}
Code used for the Bayesian inference, numerical simulations, and data analysis presented in this work is publicly available at GitHub and archived on Zenodo at Ref.~\cite{codes}. The code was developed in MATLAB R2025a (MathWorks).

\bibliography{draft.bib}
\clearpage
\onecolumngrid

\setcounter{figure}{0}
\renewcommand{\thefigure}{\arabic{figure}}
\renewcommand{\figurename}{Supplementary Figure}

\renewcommand{\theequation}{S\arabic{equation}}

\renewcommand{\thesection}{S\arabic{section}}

\section{Supplementary Note 1. Explicit expressions for the analytical PL model}
\label{app:explicit_PL}

Equation~(8) in the main text provides a compact forward model for the photoluminescence (PL) map. For reproducibility and to make the dependence on the control parameters $(B^{\mathrm{bias}},\phi)$ explicit, we give here the closed-form expressions of the resonance deviations $\delta_i(B^{\mathrm{bias}},\phi)$ entering Eq.~(8), together with the rotation conventions used to transform between laboratory and sample frames. The expressions are lengthy and are therefore collected in this Supplementary Note.

We first express the externally applied magnetic field in the laboratory frame as the sum of a transverse component confined to the $xy$-plane and an axial component along 
$z$. The transverse field has fixed magnitude $b_\perp$ and orientation set by the phase $\varphi_0$, while $b_z$ denotes the axial field. In the laboratory unit vectors $\{\mathbf{u_x},\mathbf{u_y},\mathbf{u_z}\}$, the field reads
\begin{equation} 
    \mathbf{B^{ext}}=b_\perp\left[\cos(\varphi_0)\mathbf{u_x}+\sin(\varphi_0)\mathbf{u_y}\right]+b_z\mathbf{u_z}.
\end{equation}

The total magnetic field experienced by the NV ensemble in the laboratory frame is obtained by adding the controllable axial bias $B^\text{bias}$ field to the external field defined above. This yields
\begin{equation}
     \mathbf{B^\text{lab}}=\mathbf{B^{ext}} + B^\text{bias}\mathbf{u_z} = b_\perp\left[\cos(\varphi_0)\mathbf{u_x}+\sin(\varphi_0)\mathbf{u_y}\right]+[B^\text{bias}+b_z]\mathbf{u_z}.
\end{equation}

To obtain the magnetic field in the crystal frame, we apply the rotation that maps laboratory coordinates onto the sample coordinates. With the convention introduced in the main text, the corresponding rotation operator is
\begin{equation}
    O^{\text{s}}_{\text{lab}} = R_{[111]}(-\zeta)\cdot R_{[1\bar10]}(-\beta)\cdot R_z(-\alpha),
\end{equation}
where the negative angles account for the change of basis from the laboratory frame to the sample frame. Here $R_u(\theta)$ denotes the right-handed rotation operator corresponding to a rotation by an angle $\theta$ about the unit vector $\mathbf{u}$. Rather than writing the full matrix explicitly, we use its action on an arbitrary vector $\mathbf{v}$, given by Rodrigues’ formula,
\begin{equation}
    R_{\mathbf{u}}(x)\cdot\mathbf{v} = \mathbf{v}\cos(x)+ \mathbf{u}(\mathbf{u}\cdot\mathbf{v})(1-\cos(x))-\mathbf{v}\times\mathbf{u}\sin(x),
\end{equation}
where $\mathbf{u}$ is assumed to be normalized. This expression is used throughout to rotate vectors from the laboratory frame into the sample frame and vice versa. For instance, the rotation $R_{[111]}$ corresponds to taking $\mathbf{u}=(1,1,1)/\sqrt{3}$ and $\theta=-\zeta$. In addition, during the measurement the sample is rotated by an angle  $\phi$ around the laboratory $z$-axis, which is incorporated through the corresponding rotation $R_z(-\phi)$. The magnetic field expressed in the sample frame is therefore obtained as
\begin{equation}
    \mathbf{B^\text{s}}=O_\text{lab}^\text{s}\cdot  R_z(-\phi)\cdot\mathbf{B}^\text{lab}.
\end{equation}

With that, the magnetic field components in the sample frame can be written explicitly as functions of the swept experimental parameters $(B^\text{bias},\phi)$, the external field parameters $(b_z, b_\perp, \varphi_0)$, and the three angles $(\alpha,\beta,\zeta)$ specifying the crystal orientation. Carrying out the rotations and collecting terms yields
\begin{align}
      \mathbf{B^\text{s}}&= R_{[111]}(-\zeta)\cdot R_{[1\bar10]}(-\beta)\cdot R_z(-\alpha)\cdot R_{z}(-\phi)\cdot\mathbf{B^\text{lab}} \nonumber \\
      &= R_{[111]}(-\zeta)\cdot R_{[1\bar10]}(-\beta)\cdot R_z(-\alpha)\cdot\left\{b_\perp\left[\cos(\varphi_0-\phi)\mathbf{u_x}+\sin(\varphi_0-\phi)\mathbf{u_y}\right]+[B^\text{bias}+b_z]\mathbf{u_z}\right\} \nonumber\\
     &= R_{[111]}(-\zeta)\cdot R_{[1\bar10]}(-\beta)\cdot\left\{b_\perp\left[\cos(\varphi_0-\phi-\alpha)\mathbf{u_x}+\sin(\varphi_0-\phi-\alpha)\mathbf{u_y}\right]+[B^\text{bias}+b_z]\mathbf{u_z}\right\} \nonumber\\
     &= R_{[111]}(-\zeta)\cdot \left\{\left[\frac{b_\perp}{\sqrt{2}}\left(\cos\left(\varphi_0-\phi-\alpha+\frac{\pi}{4}\right)+\cos\beta\sin\left(\varphi_0-\phi-\alpha+\frac{\pi}{4}\right)\right)+\frac{B^\text{bias}+bz}{\sqrt{2}}\sin\beta\right]\mathbf{u_x} \right. \nonumber \\
    &\phantom{= R_{[111]}(-\zeta)\cdot\big\{} + \left[\frac{b_\perp}{\sqrt{2}}\left(-\cos\left(\varphi_0-\phi-\alpha+\frac{\pi}{4}\right)+\cos\beta\sin\left(\varphi_0-\phi-\alpha+\frac{\pi}{4}\right)\right)+\frac{B^\text{bias}+bz}{\sqrt{2}}\sin\beta\right]\mathbf{u_y}\nonumber\\
    &\phantom{= R_{[111]}(-\zeta)\cdot\big\{}+\left[\left.-\frac{b_\perp}{\sqrt{2}}\sin\beta\sin\left(\varphi_0-\phi-\alpha+\frac{\pi}{4}\right)+(B^\text{bias}+b_z)\cos\beta\right]\mathbf{u_z}\right\} \nonumber \\
    &=\frac{1}{3}\left\{(B^\text{bias}+b_z)\left[\sqrt{2}\sin\beta\left(1+\sin\left(\zeta+\frac{\pi}{6}\right)\right)+\cos\beta\left(1-2\sin\left(\zeta+\frac{\pi}{6}\right)\right)\right]\right.\nonumber\\
    &\phantom{+\frac{1}{3}\big\{} +b_\perp\left[\sin\left(\varphi_0-\phi-\alpha+\frac{\pi}{4}\right)\left(\sin\beta\left(-1+2\sin\left(\zeta+\frac{\pi}{6}\right)\right)+\sqrt{2}\cos\beta\left(1+\sin\left(\zeta+\frac{\pi}{6}\right)\right)\right)\right.\nonumber\\
    &\phantom{+\frac{1}{3}\big\{ +b_\perp\big[}+\left.\left.\sqrt{6}\cos\left(\varphi_0-\phi-\alpha+\frac{\pi}{4}\right)\cos\left(\zeta+\frac{\pi}{6}\right)\right]\right\}\mathbf{u_x}\nonumber\\
    &+\frac{1}{3}\left\{(B^\text{bias}+b_z)\left[\sqrt{2}\sin\beta\left(1-\sin\left(\zeta-\frac{\pi}{6}\right)\right)+\cos\beta\left(1+2\sin\left(\zeta-\frac{\pi}{6}\right)\right)\right]\right.\nonumber\\
    &\phantom{+\frac{1}{3}\big\{} -b_\perp\left[\sin\left(\varphi_0-\phi-\alpha+\frac{\pi}{4}\right)\left(\sin\beta\left(1+2\sin\left(\zeta-\frac{\pi}{6}\right)\right)-\sqrt{2}\cos\beta\left(1-\sin\left(\zeta-\frac{\pi}{6}\right)\right)\right)\right.\nonumber\\
    &\phantom{+\frac{1}{3}\big\{ +b_\perp\big[}+\left.\left.\sqrt{6}\cos\left(\varphi_0-\phi-\alpha+\frac{\pi}{4}\right)\cos\left(\zeta-\frac{\pi}{6}\right)\right]\right\}\mathbf{u_y}\nonumber\\
    &+\frac{1}{3}\left\{(B^\text{bias}+b_z)\left[\sqrt{2}\sin\beta(1-\cos\zeta)+\cos\beta(1+2\cos\zeta)\right]\right.\nonumber\\
    &\phantom{+\frac{1}{3}\big\{}+b_\perp\left[\sin\left(\varphi_0-\phi-\alpha+\frac{\pi}{4}\right)\left(-\sin\beta(1+2\cos\zeta)+\sqrt{2}\cos(1-\cos\zeta)\right)\right.\nonumber\\
    &\phantom{+\frac{1}{3}\big\{+b_\perp\big[}+\left. \left.\sqrt{6}\cos\left(\varphi_0-\phi-\alpha+\frac{\pi}{4}\right)\sin\zeta\right]\right\}\mathbf{u_z},
\end{align}
where we have used the identity $A\sin(x)+B\cos(x)=\sqrt{A^2+B^2}\sin(x-\arctan(A/B))$ to simplify the expressions. Since the resonance deviations $\delta_i(B^{\mathrm{bias}},\phi)$ entering Eq.~(8) are defined in terms of the projections of the magnetic field onto the relevant crystallographic axes, each $\delta_i$ can be written in closed analytic form as a function of the control parameters $(B^{\mathrm{bias}},\phi)$ and of the six fixed parameters characterizing the external field and the crystal orientation. As an illustrative example, one of the nine resonance deviations can be expressed as
\begin{align}
    \delta_1(B^\text{bias},\phi) &= B_x^s-B_y^s \nonumber\\
    &= \frac{1}{3}\left[\sin\left(\zeta+\frac{\pi}{6}\right)+\sin\left(\zeta-\frac{\pi}{6}\right)\right]\left[\left(B^\text{bias}+b_z\right)\left(\sqrt{2}\sin\beta-2\cos\beta\right)\right.\nonumber\\
    &\phantom{=\frac{1}{3}\left[\sin\left(\zeta+\frac{\pi}{6}\right)+\sin\left(\zeta-\frac{\pi}{6}\right)\right]\big[}\left. +b_\perp\sin\left(\varphi_0-\phi-\alpha+\frac{\pi}{4}\right)(2\sin\beta+\sqrt{2}\cos\beta)\right]\nonumber\\
    &+\sqrt{\frac{2}{3}}b_\perp\left[\cos\left(\zeta+\frac{\pi}{6}\right)+\cos\left(\zeta-\frac{\pi}{6}\right)\right]\cos\left(\varphi_0-\phi-\alpha+\frac{\pi}{4}\right).
\end{align}
\begin{figure}[t!]
    \centering
     \includegraphics[width=0.5\linewidth]{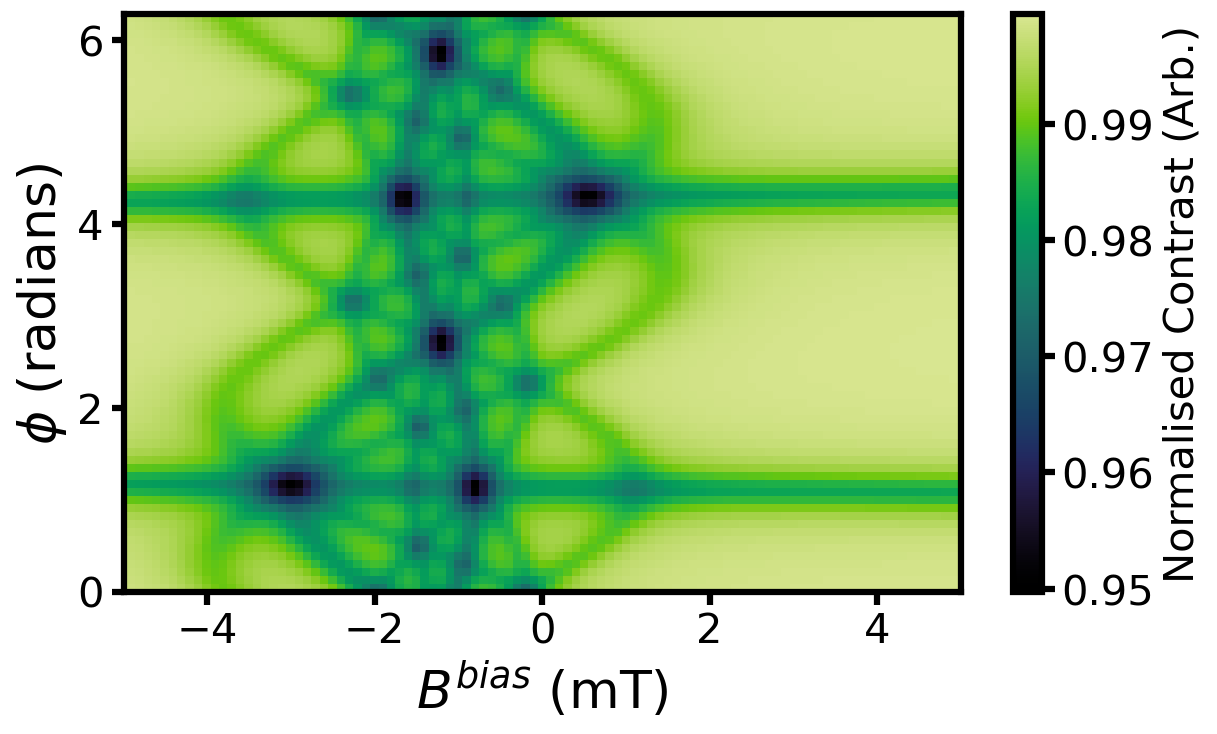}
    \caption{Simulated photoluminescence map as a function of the rotation angle $\phi$ and the bias magnetic field $B^\text{bias}$ using the inferred parameters in the magnetometry application of the main text.}
    \label{fig:experimental_magnetometry_simulation}
\end{figure}

The remaining resonance deviations are obtained from the corresponding linear combinations of the rotated field components in the sample frame. For completeness, they read
\begin{align}
\delta_2&= B_x^s-B_z^s ,\nonumber\\
\delta_3&= B_y^s-B_z^s ,\nonumber\\
\delta_4 &= B_x^s+B_y^s ,\nonumber\\
\delta_5 &= B_x^s+B_z^s ,\nonumber\\
\delta_6 &= B_y^s+B_z^s ,\nonumber\\
\delta_7 &= 2B_x^s ,\nonumber\\
\delta_8 &= 2B_y^s ,\nonumber\\
\delta_9 &= 2B_z.
\end{align}
These resonance deviations enter Eq.~(8) directly through the photoluminescence model, where the total signal is written as a superposition of lineshape functions whose arguments are the $\delta_i(B^{\mathrm{bias}},\phi)$. In our implementation, each contribution is modeled by a Lorentzian profile, so that the PL map is expressed as a sum of Lorentzian functions centered at $\delta_i=0$. In particular, we used the lineshape
\begin{equation}\label{eq:lorentzian}
    L\left[\delta_i(B^\text{bias},\phi);\Gamma\right]=\frac{1}{1+(\delta_i/\Gamma)^2},
\end{equation}
where $\Gamma$ is related to the linewidth. 

To illustrate the use of the analytical model, we show here an example of a simulated photoluminescence map generated using the parameters inferred in the magnetometry application discussed in the main text. Using the maximum a posteriori estimates of the external magnetic field parameters $(b_z,b_\perp,\varphi_0)=(1.165\, \text{mT},\;0.8081\,\text{mT},\;0.7196\,\text{rad})$ and the previously determined crystal orientation parameters $(\alpha,\,\beta,\,\zeta)=(4.7587,\;0.2342,\;0.4775)\,\text{rad}$, we compute the PL signal as a function of the control parameters ($B^\text{bias},\phi$) according to Eq.~(8), using Lorentzian profiles given by Eq.~\eqref{eq:lorentzian} with $\Gamma=0.23\,\text{mT}$ and contrast $C=0.0075$.

Supplementary figure~\ref{fig:experimental_magnetometry_simulation} shows the resulting simulated PL map. The overall structure of the map, including the position and curvature of the cross-relaxation features, is in good agreement with the experimental data presented in Fig.\ 6(a) of the main text.

\section{Supplementary Note 2. Photoluminescence normalization}
\begin{figure*}[t!]
    \centering
     \includegraphics[width=0.8\linewidth]{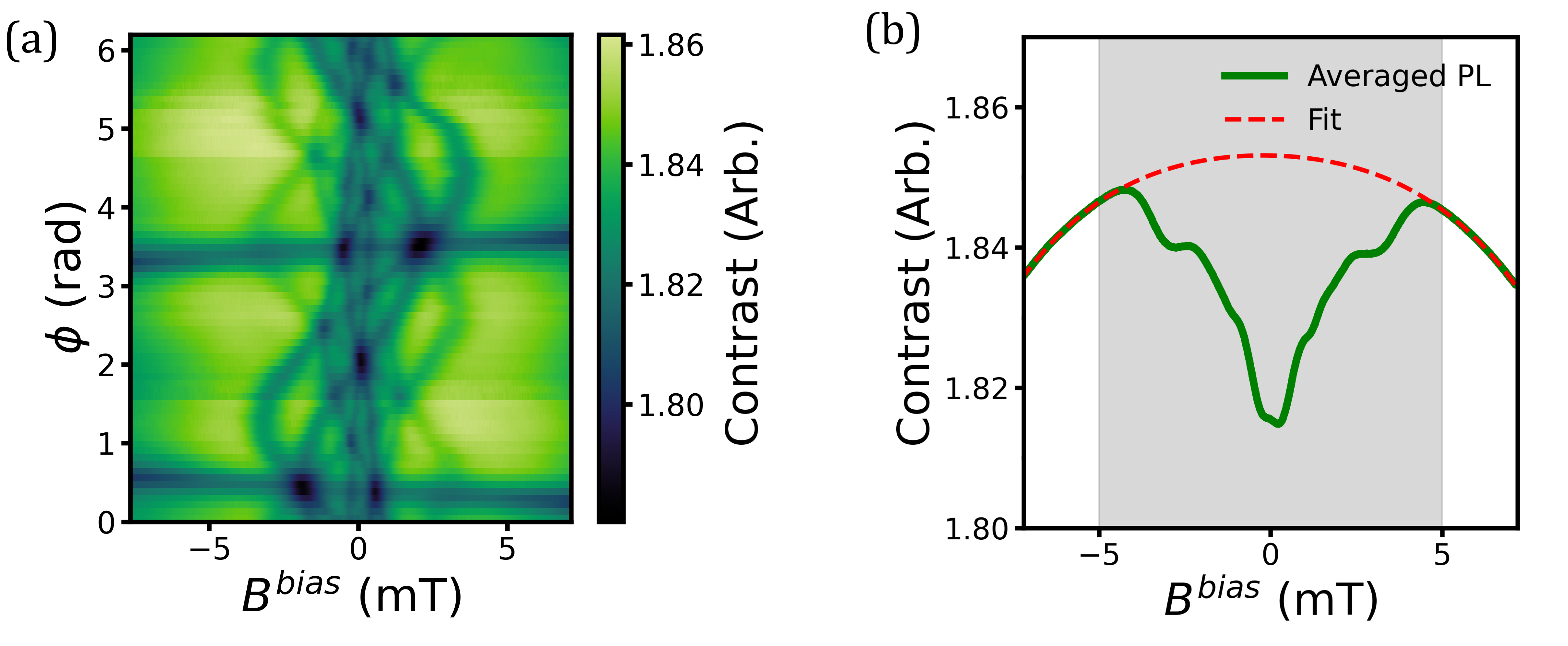}
    \caption{Photoluminescence normalization procedure. (a) Raw photoluminescence map as a function of the bias magnetic field $B^\text{bias}$ and rotation angle $\phi$. (b) PL signal averaged over all angles (green), showing the global magnetic-field-dependent background. The shaded region indicates the interval excluded from the fit to avoid the cross-relaxation features. The dashed red curve shows a fourth-order polynomial fit to the unmasked data, which is used as the baseline for normalization of the full dataset.}
    \label{fig:normalization}
\end{figure*}

In addition to the sharp features associated with cross-relaxation resonances described in the main text, the measured photoluminescence maps exhibit a smooth background dependence on the applied magnetic field magnitude. This background appears as a gradual reduction of the PL at increasing magnetic field and originates from magnetic-field-induced mixing of the spin sublevels, which modifies the spin-dependent optical cycle in NV ensembles and reduces the steady-state population in the bright $m_s=0$ state. Supplementary fig.~\ref{fig:normalization}(a) shows the original data used to determine the orientation of the diamond crystal (Fig.~4 in the main text) previous to normalization. A smooth background variation is visible, with reduced PL at large $\lvert B^\text{bias}\rvert$, superimposed on the cross-relaxation features.

To isolate the cross-relaxation structure, a global normalization of the PL signal is performed. In practice, we find that a fourth-order polynomial provides the minimal model that captures the observed curvature without overfitting, while remaining insensitive to the localized resonant features. Rather than normalizing each angular trace independently, we construct a single baseline from the full dataset. This is motivated by the presence of features that are nearly horizontal in the PL maps: a trace-by-trace normalization would partially remove or distort these features, as they would be absorbed into the fitted baseline for each individual trace. Therefore, the PL signal is first averaged over all rotation angles to obtain an effective one-dimensional dependence on the bias magnetic field $B^\text{bias}$. The PL signal averaged over $\phi$ is shown in Supplementary fig.~\ref{fig:normalization}(b). Before fitting, the central region of the magnetic-field axis containing the cross-relaxation features (shaded in gray in the figure) is masked to avoid distorting the background estimation. The normalized PL map is obtained by dividing the raw data by this global baseline.

Residual baseline variations at specific angles can introduce small systematic deviations when averaging over all angles. In practice, these effects are absorbed into the effective noise term used in the likelihood function during Bayesian inference and do not affect the extraction of the resonance positions.

\section{Supplementary Note 3. Estimate of the dynamic range }
At the low-field end, the resolution is limited by the effective linewidth $\Gamma$ and contrast of the observable PL features: if the field-induced displacements of the features become comparable to or smaller than $\Gamma$, or if several features overlap, the posterior broadens and the reconstruction becomes weakly informative. This can be further optimized by using a sample with narrower linewidth, i.e. low strain CVD diamond samples.

While the analytical framework for microwave-free cross-relaxation magnetometry does not impose a strict theoretical upper bound on the measurable magnetic field magnitude, practical limits emerge from two primary factors: (i) the maximum current and bias field-sweep windows achievable by the coil system, and (ii) the degradation of photoluminescence contrast caused by magnetically induced state mixing.

To establish a quantitative estimate for the scale of this latter effect, we consider the NV ground-state Hamiltonian
\begin{equation}
    H=DS_z^2+\gamma_e(B_\parallel S_z+B_\perp S_x),
\end{equation}
where $B_\parallel=B\cos\theta$, and $B_\perp = B\sin\theta$ are the field components parallel and perpendicular to the NV symmetry axis, $\theta$ is the angle between the applied field and that axis, $D = 2\pi \times 2.87\,\text{GHz}$ is the zero-field splitting, and $\gamma_e = 2\pi \times 2.8\,\text{MHz/G}$ is the electron gyromagnetic ratio. As the transverse field $B_\perp$ increases, it induces state mixing between the $m_s = 0$ and $m_s = \pm 1$ states. This reduces the overlap $|\langle \phi_0|0\rangle|^2$, where $|\phi_0\rangle$ is the perturbed ground state corresponding to the unperturbed $|0\rangle$ state. When this overlap reaches $1/2$, the population is split equally between the $|0\rangle$ component and the $|\pm 1\rangle$ manifold combined. At this point, optical pumping can no longer efficiently produce spin polarization, leading to a severe reduction in the photoluminescence contrast. Furthermore, analogous state mixing occurs in the triplet excited state ($^3E$).

\begin{figure*}[t!]
    \centering
     \includegraphics[width=0.4\linewidth]{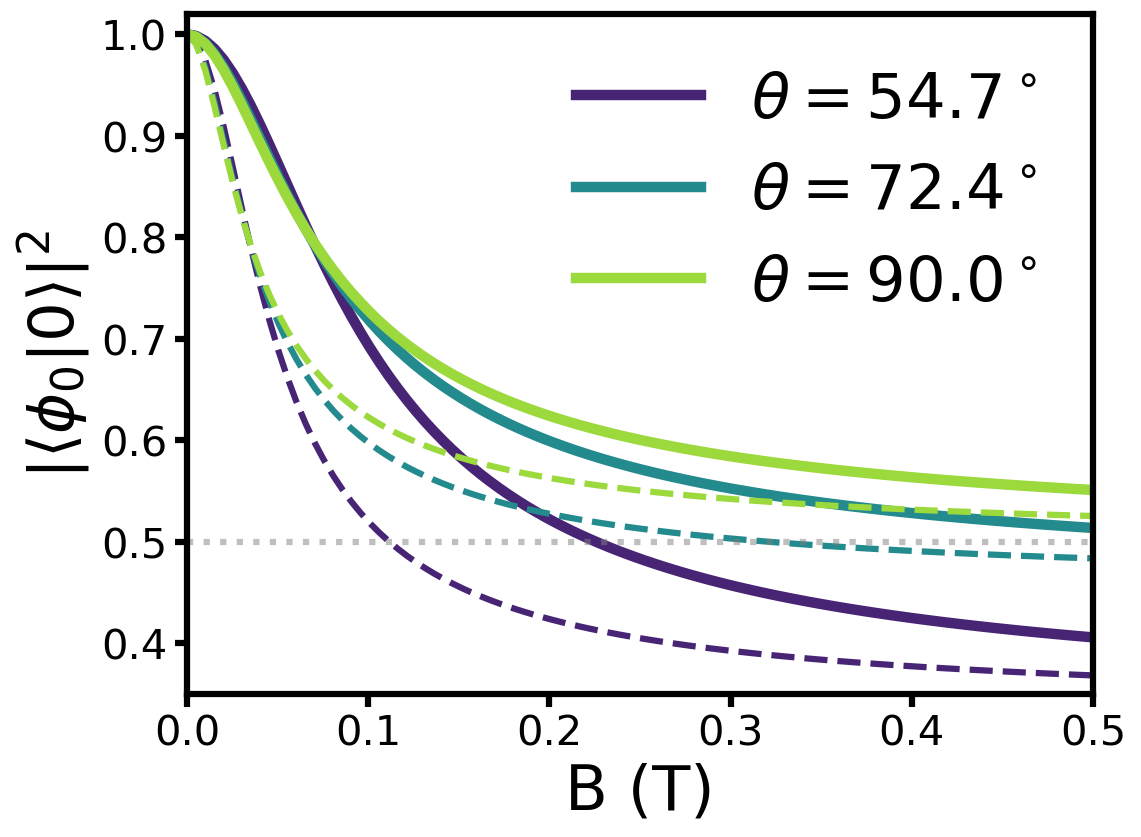}
    \caption{Magnetically induced state mixing as a function of field magnitude. Computed projection $|\langle \phi_0 | 0 \rangle|^2$ of the ground eigenstate onto the $|m_s=0\rangle$ state as a function of the magnetic field magnitude $B$. Curves represent 3 angles $\theta$ within the allowed range $[\arccos(1/\sqrt3), \pi/2]$ for cross-relaxation features. Solid lines correspond to the ground state ($D_{\mathrm{GS}} = 2.87\text{ GHz}$) and dashed lines to the triplet excited state ($D_{\mathrm{ES}} = 1.42\text{ GHz}$). The horizontal dotted line marks the $1/2$ benchmark.}
    \label{fig:optical_pumping_mixing}
\end{figure*}

Cross-relaxation features arise when the total magnetic field vector $\mathbf{B}$ yields identical projections along two distinct NV crystallographic axes. By tetrahedral symmetry, a field vector satisfying this condition is geometrically constrained such that the angle $\theta$ between $\mathbf{B}$ and each of the two respective NV axes must lie within the strict bounds $\arccos\left(\frac{1}{\sqrt{3}}\right)  \le \theta \le \frac{\pi}{2}$. In Supplementary fig.~\ref{fig:optical_pumping_mixing}, we simulate the state overlap $|\langle \phi_0 | 0 \rangle|^2$ as a function of the total field magnitude $B$ across the allowed angular range. The overlap at the lower angular limit ($\theta = \arccos\left(\frac{1}{\sqrt{3}}\right)\approx 54.7^\circ$) reaches this $1/2$ threshold at $B \approx 226\text{mT}$. In the triplet excited state, which governs the optical cycle and possesses a smaller zero-field splitting ($D_{\text{ES}} \approx 2\pi \times 1.42\text{ GHz}$), this threshold is reached at a lower field of $B \approx 112\text{ mT}$. We emphasize that this threshold does not represent an absolute upper limit beyond which all magnetometry becomes impossible. Because this severe mixing occurs at a specific, worst-case angle, other NV families will still retain sufficient overlap to produce measurable PL features. Furthermore, in an experimental setting, one could rotate the sample or the bias field relative to the crystal axes to deliberately shift the system away from these poorly polarized configurations. Therefore, the $100-200\text{ mT}$ range should be understood not as a strict physical boundary, but as a characteristic scale where state mixing begins to heavily suppress cross-relaxation features for vulnerable orientations, necessitating intentional geometric optimization to maintain a high signal-to-noise ratio.

\end{document}